\documentstyle[iopconf1]{article}
\newcommand{\Rslash}{{\not \! \!{R}}}
\begin{document}

\begin{flushright}
IFUP--TH 43/97\\ \texttt{hep-ph/9709395} \\
\end{flushright}

\title{A brief review of $R$-parity-violating couplings}

\author{Gautam Bhattacharyya
\footnote{E-mail: gautam@ibmth.difi.unipi.it ~~[{\it Invited Talk
presented at ``Beyond the Desert'', Castle Ringberg, Tegernsee,
Germany, 8-14 June 1997.}]}}

\affil{
Dipartimento di Fisica, Universit{\`a} di Pisa 
and INFN, Sezione di Pisa, I-56126 Pisa, Italy}

\beginabstract

I review the upper limits on the $R$-parity-violating ($\Rslash$)
Yukawa couplings from indirect searches.  Some limits have been
updated using recent data.

\endabstract

\section{Introduction}

In supersymmetric theories `$R$-parity' is a discrete symmetry under
which all Standard Model (SM) particles are even while their
superpartners are odd. It is defined as $R = (-1)^{(3B+L+2S)}$, where
$S$ is the spin, $B$ is the baryon-number and $L$ is the lepton-number
of the particle \cite{def_rpar,intro_rpar}. An exact $R$ implies that
superparticles could be produced only in pairs and the lightest
supersymmetric particle (LSP) is stable. However, $B$- and
$L$-conservations are not ensured by gauge invariance and therefore it
is worthwhile to investigate what happens when $R$-parity is
violated. In this talk, I concentrate on explicit $R$-parity violation
\cite{explicit}. Notice that in the Minimal Supersymmetric Standard
Model (MSSM), the gauge quantum numbers of the Higgs superfield $H_d$
(responsible for the generation of down-type quark masses) are the
same as those of the $SU(2)$-doublet lepton superfield.  So if $L$ is
not a good quantum number, the latter can replace the former in the
Yukawa superpotential. If $B$-conservation is not assumed, no
theoretical consideration prevents one from constructing a term
involving three $SU(2)$-singlet quark superfields.  These give rise to
a $\Rslash$ superpotential:
\begin{equation}
    {\cal W}_{\not R}  =  {1\over 2}\lambda_{ijk} L_i L_j E^c_k
                        +  \lambda'_{ijk} L_i Q_j D^c_k 
                        +  {1\over 2}\lambda''_{ijk} U^c_i D^c_j D^c_k  
                        + \mu_i L_i H_u,
\label{R-parity}
\end{equation}
where $L_i$ and $Q_i $ are SU(2)-doublet lepton and quark superfields
respectively; $E^c_i, U^c_i, D^c_i$ are SU(2)-singlet charged lepton,
up- and down-quark superfields respectively; $H_u$ is the Higgs
superfield which is responsible for the generation of up-type quark
masses; $\lambda_{ijk}$- and $\lambda'_{ijk}$-types are $L$-violating
while $\lambda''_{ijk}$-types are $B$-violating Yukawa couplings.
$\lambda_{ijk}$ is antisymmetric under the interchange of the first
two generation indices, while $\lambda''_{ijk}$ is antisymmetric under
the interchange of the last two. Thus there could be 27
$\lambda'$-type and 9 each of $\lambda$- and $\lambda''$-type
couplings. Hence including the 3 additional bilinear $\mu$-terms
($\mu_i$), there are 48 additional parameters in the theory.

\section{Indirect limits}

\subsection{Proton stability}
Non-observation of proton decay places very strong bounds on the
simultaneous presence of both $\lambda'$ and $\lambda''$ couplings.  The
combinations involving the lighter generations are most tightly
constrained: $\lambda'_{11k} \lambda''_{11k} \leq 10^{-22}$ for $k =
2, 3$ and for $\tilde{m} = 100$ GeV \cite{HK}.  Detailed analyses have
been presented in \cite{sher,vissani}.  It has been shown in
\cite{vissani} that any flavour combination of the product
$\lambda'.\lambda'' \le 10^{-10}$ for $\tilde{m} = 100$ GeV.

\subsection{$n$--$\bar{n}$ oscillation}
Goity and Sher \cite{goity} have put a (model independent) limit
$\lambda''_{113} \le 10^{-4}-10^{-5}$ for $m_{\tilde{t}} = 100$ GeV
from the consideration of electroweak box graph induced $n$--$\bar{n}$
oscillation. The corresponding limit on $\lambda''_{112}$ is diluted by
a relative factor of $m_s^2/m_b^2$.  However the best constraint on
$\lambda''_{112}$ comes from the consideration of double nucleon decay
into two kaons and the bound is estimated to be $\le 10^{-6}-10^{-7}$
\cite{goity}.

\subsection{$\nu_e$-Majorana mass}
An approximate expression for $\nu_e$-Majorana mass induced by an
appropriate $\lambda$ (or $\lambda'$), {\it via} self-energy type
diagrams, is  
\begin{equation}
\delta m_{\nu_e} \approx {\lambda^2 N_c\over {8\pi^2}} 
                      {1\over {\tilde{m}^2}} M_{\rm SUSY} m^2.
\label{nu_mass}
\end{equation}
In the numerator of the RHS of eq.~(\ref{nu_mass}), one power of $m$
(the fermion mass in the loop) appears due to chirality-flip in an
internal line. The left-right sfermion mixing has been assumed to be $
M_{\rm SUSY} m$.  Requiring $\delta m_{\nu_e} \le 5$ eV and assuming
$M_{\rm SUSY}=\tilde{m}$, the $\lambda_{133}$-induced interaction with
$\tau\tilde{\tau}$ loops ($N_c = 1$) yields the constraint ($1\sigma$)
$\lambda_{133} \le 0.003$ for $m_{\tilde{\tau}} = 100$ GeV
\cite{dimo}. The $\lambda'_{133}$-induced diagrams with $b\tilde{b}$
loops ($N_c = 3$) leads to $\lambda'_{133} \le 0.0007$ for
$m_{\tilde{b}} = 100$ GeV \cite{grt}.

\subsection{Neutrinoless double beta decay}
It has been known for a long time that neutrinoless double beta decay
($(\beta\beta)_{0\nu}$) is a sensitive probe of $L$-violating
processes. In $\Rslash$ scenario, the process $dd \rightarrow
uue^{-}e^{-}$ is mediated by $\tilde{e}$ and $\tilde{N}$ (neutralino)
or by $\tilde{q}$ and $\tilde{g}$, yielding $\lambda'_{111} \le
0.00035$ for a squark and gluino mass of 100 GeV
\cite{rabi_db,klap}. The particular combinations of nuclear matrix
elements that lead to bounds on $\lambda'_{111}$ do not significantly
suffer from the uncertainties of model approximations in those
calculations \cite{klap}.  A bound on the product
coupling $\lambda'_{113} \lambda'_{131} \le 3 \times 10^{-8}$ has been
placed from the consideration of the diagrams involving the exchange
of one $W$ boson and one scalar boson \cite{babu_db}.

\subsection{Charged-current universality}
Universality of the lepton and quark couplings to the $W$-boson is
violated by the presence of $\lambda$- and $\lambda'$-type couplings.
The scalar-mediated new interactions could be written in the same
$(V-A) \otimes (V-A)$ structure as the $W$-exchanged SM graph. The
experimental value of $V_{ud}$ is related to $V_{ud}^{\rm SM}$ by
\begin{equation}
|V_{ud}^{\rm exp}|^2 \simeq |V_{ud}^{\rm SM}|^2 \left[1+{{2 r'_{11k}
(\tilde{d}_R^k)}\over {V_{ud}}} - 2 r_{12k}(\tilde{e}_R^k)\right],
\end{equation}
where, 
\begin{equation}
r_{ijk}(\tilde{l}) = (M_W^2/g^2)(\lambda_{ijk}^2/m_{\tilde{l}}^2),  
\end{equation}
and $r'_{ijk}$ is defined using $\lambda'_{ijk}$ analogously. Assuming
the presence of only one $\Rslash$ coupling at a time, one obtains,
for a common $\tilde{m} = 100$ GeV, $\lambda_{12k} \le 0.05~(1\sigma)$
and $\lambda'_{11k} \le 0.02~(2\sigma)$, for each $k$
\cite{bgh,dreiner}.

\subsection{$e$--$\mu$--$\tau$ universality}
In the presence of $\lambda'$-type interaction, the ratio $R_{\pi}
\equiv \Gamma(\pi \rightarrow e\nu)/ \Gamma(\pi \rightarrow \mu\nu)$
takes the form
\begin{equation}
R_\pi = R_\pi^{\rm SM} \left[1+ {2\over V_{ud}}\left\{r'_{11k}
(\tilde{d}_R^k) - r'_{21k}(\tilde{d}_R^k)\right\}\right].
\end{equation}
A comparison with experimental results yields, for a common mass
$\tilde{m} = 100$ GeV and at $1\sigma$, $\lambda'_{11k} \le 0.05$ and
$\lambda'_{21k} \le 0.09$, for each $k$, assuming only one coupling at
a time \cite{bgh}.

Similarly, from the consideration of $\Gamma(\tau
\rightarrow e\nu\bar{\nu})/ \Gamma(\tau \rightarrow \mu\nu\bar{\nu})$,
one obtains, $\lambda_{13k} \le 0.06$ and $\lambda_{23k} \le 0.06$,
for each $k$, at $1\sigma$ and for $\tilde{m} = 100$ GeV
\cite{bgh,dreiner}.

\subsection{$\nu_\mu$--$e$ scattering}
The neutrino-electron scattering cross section at low energies are 
given by 
\begin{eqnarray}
\sigma(\nu_\mu e) & = & {G_F^2 s\over{\pi}}(g_L^2 + {1\over 3} g_R^2),
                                         \nonumber \\
\sigma(\bar{\nu}_\mu e) & = & {G_F^2 s\over{\pi}}
({1\over 3}g_L^2 +  g_R^2); 
\end{eqnarray}
where in the presence of $\Rslash$ interactions 
$(x_W \equiv \sin^2\theta_W)$
\begin{eqnarray}
g_L & = & x_W - {1\over 2} - ({1\over 2} + x_W) r_{12k}(\tilde{e}_R^k),
                                     \nonumber \\
g_R & = & x_W + r_{121}(\tilde{e}_L^1) + r_{231}(\tilde{e}_L^3) 
     -   x_W r_{12k}(\tilde{e}_R^k). 
\end{eqnarray}
The upper limits (at $1\sigma$) are $\lambda_{12k} \le 0.34$, 
$\lambda_{121} \le 0.29$ and $\lambda_{231} \le 0.26$ for $\tilde{m} 
= 100$ GeV \cite{bgh}.

\subsection{Atomic parity violation (APV)}
The parity-violating part of the Hamiltonian of the electron-hadron 
interaction is 
\begin{equation}
H = {G_F\over{\sqrt{2}}} \left(C_{1i} \bar{e}\gamma_\mu\gamma_5 e
\bar{q}_i\gamma_\mu q_i + C_{2i} \bar{e}\gamma_\mu e
\bar{q}_i\gamma_\mu \gamma_5 q_i\right), 
\end{equation}
where  $i$ runs over the $u$- and $d$-quarks. The $\Rslash$
interactions modify $C_{1i}$ and $C_{2i}$ in the following way:
\begin{eqnarray}
 C_{1u} & = &  -{1\over 2} + {4\over 3} x_W - r'_{11k}(\tilde{d}_R^k) 
+ ({1\over 2}  - {4\over 3} x_W) r_{12k}(\tilde{e}_R^k), \nonumber \\
 C_{2u} & = & -{1\over 2} + 2 x_W - r'_{11k}(\tilde{d}_R^k) 
+ ({1\over 2} - 2 x_W) r_{12k}(\tilde{e}_R^k), \nonumber \\
 C_{1d} & = & {1\over 2} - {2\over 3} x_W +  r'_{1j1}(\tilde{q}_L^j) 
- ({1\over 2}  - {2\over 3} x_W) r_{12k}(\tilde{e}_R^k),  \\
 C_{2d} & = & {1\over 2} - 2 x_W - r'_{1j1}(\tilde{q}_L^j) 
- ({1\over 2} - 2 x_W) r_{12k}(\tilde{e}_R^k). \nonumber
\end{eqnarray}
Using the SM value of the weak charge $Q_W^{\rm SM} = -73.17 \pm
0.13$ \cite{mar} and the new experimental number $Q_W^{\rm exp} =
-72.11 \pm 0.93$ \cite{wood}, the upper limits on the $\lambda'_{1j1}$
couplings can be significantly improved as $\le 0.035$ for $\tilde{m}
= 100$ GeV \cite{dreiner,DBC}. Note that these couplings are relevant for
the `$\Rslash$ squark explanation' of the recent large-$Q^2$ HERA anomaly
\cite{altarelli}. 

\subsection{$\nu_\mu$ deep-inelastic scattering}
The left- and the right-handed couplings of the $d$-quark in neutrino 
interactions are modified by the $\Rslash$ couplings as 
\begin{eqnarray} 
g_L^d & = &(-{1\over 2} + {1\over 3} x_W)(1-r_{12k} (\tilde{e}_R^k))
 - r'_{21k} (\tilde{d}_R^k), \nonumber \\
g_R^d & = & {1\over 3} x_W + r'_{2j1} (\tilde{d}_L^j)
            -{1\over 3} x_W r_{12k} (\tilde{e}_R^k). 
\end{eqnarray}
The bounds, for $\tilde{m} = 100$ GeV, are 
$\lambda'_{21k} \le 0.11$ ($1\sigma$) and 
$\lambda'_{2j1} \le 0.22$ ($2\sigma$) \cite{bgh}. 

\subsection{Quark mixing: $K^{+} \rightarrow \pi^+ \nu\bar{\nu}$ or 
$D^0$--$\bar{D}^0$ mixing} Consideration of only one non-zero
$\Rslash$ coupling with indices related to the weak basis of fermions
generates more than one non-zero coupling with different flavour
structure in the mass basis. However, all what we know about quark
mixing is the relative rotation between the left-handed up and down
sectors given by the CKM matrix ($V_{\rm CKM} = U_L^{u\dag}
U_L^d$). The absolute mixing in either sector is not known. If we
assume $V_{\rm CKM} = U_L^d$, the strongest bounds come from $K^{+}
\rightarrow \pi^+ \nu\bar{\nu}$ which, in the presence of
$\Rslash$-interactions, proceeds at tree level. The bounds are
$\lambda'_{ijk} \le 0.012$ (90\% CL), for $m_{\tilde{d}_R^k} = 100$
GeV and for $j = 1$ and $2$ \cite{agashe}. If we assume the other
extreme, {\it i.e.} $V_{\rm CKM} = U_L^{u}$, the bounds from $K^+$
decay become invalid. The best bounds in this case arise from
$D^0$--$\bar{D}^0$ mixing and the upper limits are considerably
relaxed becoming 0.20 \cite{dreiner,agashe}. Although the latter is a
much more conservative estimate than the former, all of them are
nevertheless basis-dependent bounds.

\subsection{$\tau$-decays}
The decay $\tau^{-} \rightarrow \bar{u}d\nu_\tau$ proceeds in the SM
through a tree-level $W$-exchanged graph. The scalar-exchanged graph
induced by $\lambda'_{31k}$ can be written in the same $(V-A)\otimes
(V-A)$ form by a Fierz rearrangement. Using the experimental input
\cite{pdg}:
\begin{equation}
   Br(\tau^- \rightarrow \pi^- \nu_\tau)  =  0.113 \pm 0.0015, 
\end{equation}
with $f_{\pi^{-}} = (130.7 \pm 0.1 \pm 0.36) \; {\rm MeV}$,
one obtains $\lambda'_{31k}  \le  0.10$ ($1\sigma$) for  
$m_{\tilde{d}_R^k} = 100$ GeV \cite{dtau} (I have updated this bound).

\subsection{$D$-decays}  
The tree-level process $c \rightarrow s e^+ \nu_e$ is mediated 
by a $W$ exchange in the SM and by a scalar boson exchange in 
$\lambda'$-induced interaction. By a Fierz transformation it is 
possible to express the latter in the same $(V-A)\otimes (V-A)$ form
as the former. Using the experimental input \cite{pdg}:
\begin{equation}
 \frac{ Br(D^+ \rightarrow \bar{K}^{0 \ast} \mu^+ \nu_\mu ) }
        { Br(D^+ \rightarrow \bar{K}^{0 \ast} e^+ \nu_e ) }
     =  0.94 \pm 0.16, 
\end{equation}
one obtains (at $1\sigma$) $\lambda'_{12k} \le  0.29$ and 
$\lambda'_{22k} \le  0.18$, for $\tilde{m} = 100$ GeV   
\cite{dtau}. The form factors related to 
the hadronic matrix elements cancel in the ratios, thus
making the prediction free from the large theoretical
uncertainties associated with those matrix elements.

\subsection{LEP precision measurements}
Heavy virtual chiral fermions induce sizable loop corrections to
$\Gamma (Z\rightarrow f\bar{f})$ ($f$ is a light fermion) {\it via}
fermion-sfermion mediated triangle graphs. Since vertices involving
$\lambda'_{i3k}$ \cite{bes} or $\lambda''_{3jk}$ \cite{bcs} could
allow top quark in internal lines of a triangle diagram, the bounds on
them are most interesting.  For $\tilde{m} = 100$ GeV and at
$1\sigma$, the following bounds emerge ($R_l = \Gamma_{\rm
had}/\Gamma_l$; $R_l^{\rm SM} = 20.756$ with $m_H$ treated as a free
parameter):
\begin{eqnarray}
\lambda'_{13k} & \le & 0.34 \leftarrow
R_e^{\rm exp} = 20.757 \pm 0.056, \nonumber \\
\lambda'_{23k} & \le & 0.36 \leftarrow
R_\mu^{\rm exp} = 20.783 \pm 0.037, \nonumber \\
\lambda'_{33k} & \le & 0.48 \leftarrow
R_\tau^{\rm exp} = 20.823 \pm 0.050, \\
\lambda''_{3jk} & \le & 0.50 \leftarrow
R_l^{\rm exp} = 20.775 \pm 0.027. \nonumber
\end{eqnarray}
I have updated these limits\footnote{While extracting limits on
$\lambda''$, leptonic universality in $R_l$ is assumed since
$\lambda''$-Yukawa couplings do not directly couple to any leptonic
flavour.} using the most recent experimental numbers for the LEP
observables presented at the EPS meeting at Jerusalem \cite{lep}.

\section{Summary} 
To summarise, I have presented, in Table 1 and Table 2, the best
indirect upper bounds to date on the 45 $\Rslash$ Yukawa couplings and
the processes from which they are constrained. I also present, in
Table 3, some important product couplings -- their upper limits and
the processes that constrain them \cite{CR,barbieri}.  The limits on
$\lambda''_{123}$, $\lambda''_{212}$, $\lambda''_{213}$ and
$\lambda''_{223}$, presented in Table 1, correspond to the requirement
that these couplings remain perturbative upto the GUT scale
\cite{biswa}. In this short review, I have basically followed, with
some modifications, the style of my earlier review on
$R$-parity-violating couplings \cite{susy96} but updated quite a few
limits.

\section*{Acknowledgements} 
I thank the organizers of `Beyond the Desert: Accelerator- and
Non-accelerator Approaches' at Castle Ringberg for arranging a
successful meeting. I also thank D. Choudhury, J. Ellis,
A. Raychaudhuri and K. Sridhar for enjoyable collaborations on
$R$-parity violation and H. Dreiner, J. Kalinowski and Y. Sirois for
discussions.

\begin{table}
\begin{center}
\footnotesize\rm 
\caption{Upper limits ($1\sigma$) on $\lambda$- and
$\lambda''$-couplings for $\tilde{m} = 100$ GeV. The numbers with (*)
correspond to $2\sigma$ limits and those with (\dag) are not
phenomenological limits.}
\begin{tabular}{llllll}
\topline
$ijk$ & $\lambda_{ijk}$ & Sources & $ijk$ & $\lambda''_{ijk}$ & Sources \\
\midline
121 & 0.05(*) & CC univ. & 112 & $10^{-6}$ & 
Double nucleon decay \\
122 & 0.05(*) & CC univ. & 113 & $10^{-4}$ & $n$--$\bar{n}$
osc. \\
123 & 0.05(*) & CC univ. & 123 & 1.25(\dag) & Pert. unitarity \\
131 & 0.06 & 
$\Gamma(\tau
\rightarrow e\nu\bar{\nu})/ \Gamma(\tau \rightarrow \mu\nu\bar{\nu})$
& 212 & 1.25(\dag) & Pert. unitarity \\
132 & 0.06 & 
$\Gamma(\tau
\rightarrow e\nu\bar{\nu})/ \Gamma(\tau \rightarrow \mu\nu\bar{\nu})$
& 213 & 1.25(\dag) & Pert. unitarity \\
133 & 0.003 & $\nu_e$- mass & 223 & 1.25(\dag) & Pert. unitarity \\
231 & 0.06 & 
$\Gamma(\tau
\rightarrow e\nu\bar{\nu})/ \Gamma(\tau \rightarrow \mu\nu\bar{\nu})$
& 312 & 0.50 & $R_l$ (LEP1) \\
232 & 0.06 & 
$\Gamma(\tau
\rightarrow e\nu\bar{\nu})/ \Gamma(\tau \rightarrow \mu\nu\bar{\nu})$
& 313 & 0.50 & $R_l$ (LEP1) \\
233 & 0.06 & 
$\Gamma(\tau
\rightarrow e\nu\bar{\nu})/ \Gamma(\tau \rightarrow \mu\nu\bar{\nu})$
& 323 & 0.50 & $R_l$ (LEP1) \\
\bottomline
\end{tabular}
\end{center}
\end{table}

\begin{table} 
\begin{center}
\footnotesize\rm 
\caption{Upper limits ($1\sigma$) on $\lambda'$-couplings for
$\tilde{m} = 100$ GeV. The numbers with (*) correspond to $2\sigma$
limits and those with (\ddag) are basis-dependent limits.}
\begin{tabular}{lllllllll}
\topline
$ijk$ & $\lambda'_{ijk}$ & Sources & $ijk$ & $\lambda'_{ijk}$ &
Sources & $ijk$ & $\lambda'_{ijk}$ & Sources \\
\midline
111 & 0.00035 & $(\beta\beta)_{0\nu}$ & 211 & 0.09 & 
$R_\pi$ ($\pi$-decay) & 
311 & 0.10 & 
$\tau^{-} \rightarrow \pi^{-}\nu_\tau$ \\ 
112 & 0.02(*) & CC univ. & 212 & 0.09 &  
$R_\pi$ ($\pi$-decay) & 
312 & 0.10 & 
$\tau^{-} \rightarrow \pi^{-}\nu_\tau$ \\
113 & 0.02(*) & CC univ. & 213 & 0.09 &  
$R_\pi$ ($\pi$-decay) & 
313 & 0.10 & 
$\tau^{-} \rightarrow \pi^{-}\nu_\tau$ \\
121 & 0.035(*) & APV & 221 & 0.18 & $D$-decay &
321 & 0.20(\ddag) & $D^0$--$\bar{D}^0$ mix. \\
122 & 0.02 & $\nu_e$-mass  & 222 & 0.18 & $D$-decay &
322 & 0.20(\ddag) & $D^0$--$\bar{D}^0$ mix. \\
123 & 0.20(\ddag) & $D^0$--$\bar{D}^0$ mix. & 223 & 0.18 & $D$-decay &
323 & 0.20(\ddag) & $D^0$--$\bar{D}^0$ mix. \\
131 & 0.035(*) & APV & 231 & 0.22(*) & $\nu_\mu$ {\it d.i} scatter. & 
331 & 0.48 & $R_\tau$(LEP) \\
132 & 0.34 & $R_e$ (LEP) & 232 & 0.36 & $R_\mu$ (LEP) & 332 & 
0.48 & $R_\tau$(LEP) \\
133 & 0.0007 & $\nu_e$-mass & 233 & 0.36 & $R_\mu$ (LEP) & 333 & 
0.48 & $R_\tau$(LEP) \\
\bottomline
\end{tabular}
\end{center}
\end{table}

\begin{table} 
\begin{center}
\footnotesize\rm 
\caption{Upper limits on some important product couplings for
$\tilde{m} = 100$ GeV.}
\begin{tabular}{llllll}
\topline
Combinations & Limits & Sources & Combinations & Limits & Sources \\
\midline
$\lambda'_{11k} \lambda''_{11k}$ & $10^{-22}$ & Proton decay &
$\lambda'_{ijk} \lambda''_{lmn}$ & $10^{-10}$ & Proton decay \\
$\lambda_{1j1} \lambda_{1j2}$ & $7. 10^{-7}$ & $\mu \rightarrow 3e$ & 
$\lambda_{231} \lambda_{131}$ & $7. 10^{-7}$ & $\mu \rightarrow 3e$ \\
Im $\lambda'_{i12} \lambda'^{*}_{i21}$ & $8. 10^{-12}$ & $\epsilon_K$ & 
$\lambda'_{i12} \lambda'_{i21}$ & $1. 10^{-9}$ & $\Delta m_K$ \\ 
$\lambda'_{i13} \lambda'_{i31}$ & $8. 10^{-8}$ & $\Delta m_B$ &
$\lambda'_{1k1} \lambda'_{2k2}$ & $8. 10^{-7}$ & $K_L \rightarrow \mu
e$ \\
$\lambda'_{1k1} \lambda'_{2k1}$ & $5. 10^{-8}$ & $\mu {\rm Ti} 
\rightarrow e {\rm Ti}$ & 
$\lambda'_{11j} \lambda'_{21j}$ & $5. 10^{-8}$ & $\mu {\rm Ti} 
\rightarrow e {\rm Ti}$ \\
\bottomline
\end{tabular}
\end{center}
\end{table}

\end{document}